# Fabrication and properties of L-arginine-doped PCL electrospun composite scaffolds

S.I. Goreninskii[1], E.N. Bolbasov[1], E.A. Sudarev[1], K.S. Stankevich[1], Y.G. Anissimov[2,3], A.S. Golovkin[4], A. I. Mishanin[4], A. N. Viknianshchuk[5], V.D. Filimonov[1], S.I. Tverdokhlebov[1]*

1. Tomsk Polytechnic University, Tomsk, Russian Federation.

2. Griffith University, School of Natural Sciences, Nathan, Queensland, Australia.

3. Queensland Micro- and Nanotechnology Centre, Griffith University, Nathan, Queensland, Australia.

4. Almazov National Medical Research Centre, St. Petersburg, Russian Federation.

5. Saint Petersburg University, St. Petersburg, Russian Federation.

Corresponding author: Sergei I. Tverdokhlebov, e-mail: tverd@tpu.ru, phone: +7(3822)56-34-37, Tomsk Polytechnic University, 30 Lenin Avenue, Tomsk, 634050, Russian Federation.

**Abstract**

The article describes fabrication and properties of composite fibrous scaffolds obtained by electrospinning of the solution of poly(ε-caprolactone) and arginine in common solvent. The influence of arginine content on structure, mechanical, surface and biological properties of the scaffolds was investigated. It was found that with an increase of arginine concentration diameter of the scaffold fibers was reduced, which was accompanied by an increase of scaffold strength and Young modulus. It was demonstrated that porosity and water contact angle of the scaffold are independent from arginine content. The best cell adhesion and viability was shown on scaffolds with arginine concentration from 0.5 to 1 % wt.

1. **Introduction**

Electrospun fibrous scaffolds are common material for cardiovascular grafts [1]. Owing to high biocompatibility and good mechanical properties, poly(ε-caprolactone) is the one of the most widely used polymers for these devices [2]. The main disadvantages of PCL cardiovascular grafts are high risk of thrombosis and low endothelialisation rate [3]. To overcome these drawbacks, a number of methods of surface and bulk modification of PCL grafts were developed [4].

Application of L-arginine for modification of cardiovascular implants is promising as L-arginine is natural substrate for a family of enzymes called NO-synthases. By NO-synthases L-arginine is being metabolized to citrulline and nitric oxide (II) [5]. With that, nitric oxide (II) is one of the most important substances in cardiovascular system, sustaining vascular tone and blood pressure, inhibiting monocyte and neutrophils adhesion to endothelium, controlling smooth muscle cells proliferation and stimulating wound healing [6]. It is known that L-arginine immobilization on the surface of polymer films decreases their thrombogenicity by reducing blood calcification rate, thus improving their hemocompatibility [7,8]. The disadvantage of the surface immobilization of L-arginine is reduce of pharmacological activity during graft degradation and substitution by self-tissues.

Thus, the development of biodegradable L-arginine-doped polymer composites (in particular, PCL-based) is a relevant task. According to our knowledge, such composites and fibrous scaffolds made of them have not been produced yet. The most significant limitation concerning this problem is the absence of common solvents for PCL and L-arginine. By our group it was found that hexafluoro-2-propanol dissolves PCL as well as L-arginine giving an opportunity of developing novel composite materials. The aim of the present work is to obtain new composite materials based on PCL and L-arginine using hexafluoro-2-propanol as a common solvent and to fabricate electrospun fibrous composite scaffolds.

## 2. Materials and methods

For scaffold fabrication poly(ε-caprolactone) $M_w$=80 kDa, (PCL, Sigma-Aldrich), L-arginine (≥99%, Sigma-Aldrich) and hexafluoro-2-propanol (≥99%, Sigma-Aldrich) as a solvent were used. Six groups of materials were formed during the study depending on L-arginine concentration in composite: 0, 0.1, 0.5, 1, 3, 7 % wt.

Composite scaffolds were fabricated using NANON-01 installation (MECC Co., Japan) at the following parameters: voltage – 20 kV, feed rate – 2 ml/h, 18G syringe tip, distance between syringe tip and collector – 150 mm. The collector was presented as a steel cylinder with length of 200 mm and 5 mm diameter. Thickness of the fabricated scaffolds was 137±12 μm.

Grafts morphology was investigated by means of scanning electron microscopy (JCM-6000Plus, Jeol). The samples were pre-coated with gold (SmartCoater, Jeol). Obtained SEM-images were analyzed by using Image J software (National Institutes of Health, USA). Water contact angle of the grafts surface was measured by sitting drop method (EasyDrop, Kruss). Arginine on the grafts surface was visualized by reaction with 1% ninhydrin solution in ethanol [9]. Thermograms of the obtained materials were recorded in the range of 0-120 degrees using DSC 204F1 Phoenix equipment (NETZSCH, Germany). Degree of crystallinity was measured using the following formula: $X_c = \frac{\Delta H_m}{\Delta H_m^0} \times 100\%$, where $X_c$ is degree of crystallinity, %; $\Delta H_m^0$ is melting enthalpy of 100% crystalline polymer (135.5 J/g for PCL [10]). Mechanical properties of the grafts were studied under uniaxial extension using Instron 3369 installation with 50N loading cell (model 2519-102, Instron) at a loading velocity of 10 mm/min. The porosity of the scaffolds was measured at room temperature by using the liquid intrusion method [11]. Statistical analysis of the obtained data was conducted by means of GraphPad Prism 6 software (GraphPad Software, Inc.) using non-parametric Kruskall-Wallis test with 0.05 level of significance.

Adhesion and viability of multipotent mesenchymal stem cells (MMSC) on fabricated materials were studied as previously described [12]. Statistical analysis was performed using Mann-Whitney ANOVA tests. Results are presented as Mean±SD.

## 3. Results and discussion

SEM-images, average fiber diameter, its distribution and water contact angles of the fabricated PCL/L-arginine grafts are shown in figure 1. Grafts formed from PCL without L-arginine addition are presented by fibers with bimodal distribution of diameter (Fig. 1a).

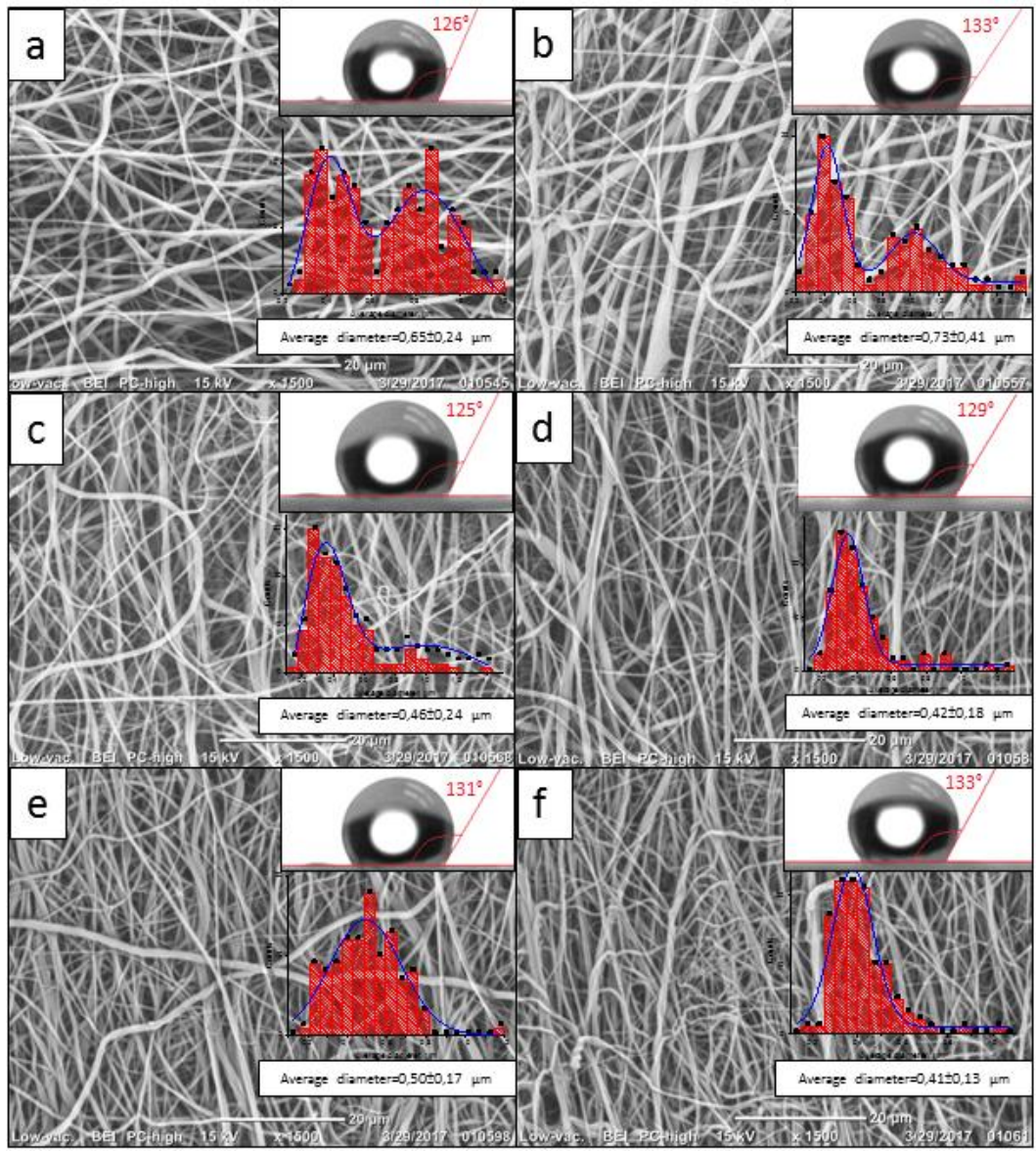

Figure 1. SEM-images, fiber diameter distribution histograms and water contact angles of PCL/L-arginine grafts depending on L-arginine content (0, 0.1, 0.5, 1, 3, 7 % wt., respectively).

L-arginine addition to the polymer solution leads to changes in structure of the scaffold fibers. With an increase of L-arginine content, the amount of fibers with a diameter in a range of 0.6-1.2 μm was reduced, as shown by peak disappearance on the plotted histograms. At concentrations of L-arginine higher than 1% fibers diameter distribution became unimodal.

Such changes in fibers structure may be explained as follows. As hexafluoro-2-propanol demonstrates relatively high acidity (pKa 9.3) and L-arginine is strong basic, reversible interactions between L-arginine and hexafluoro-2-propanol may be expected resulting in formation of such ions and zwitter-ions in polymer solution as $(CF_3)_2CHO^-$, $H_3N^+C(=NH)NH(CH_2)_3CH(NH_2)COOH$, $H_2NC(=NH_2^+)(CH_2)_3CH(NH_2)COOH$, $H_2N(C=NH)(CH_2)_3CH(NH_3^+)COOH$, $H_2N(C=NH)(CH_2)_3CH(NH_3^+)COO^-$ . Presence of these ions and zwitter-ions in the solution rises its electrical conductivity with an increase of L-arginine content. It is known that increased electrical conductivity of the spinning solution leads to enhanced splitting of its jet between capillary and collector. Formation of secondary jets takes place until the equilibrium between electrostatic forces splitting the jet and surface tension limiting that process was achieved. Thus, the possibility of secondary jets splitting increases resulting in formation of submicron fibers with unimodal diameter distribution [13].

The changes in fibers diameter distribution with L-arginine concentration had no influence on porosity of the composite scaffolds that was maintained at around 89% for all the studied groups. Increased L-arginine content also had no effect on surface wettability of the scaffolds despite hydrophilic nature of L-arginine. On the one hand, the absence of significant changes in surface wettability may be explained by distribution of L-arginine mainly inside the scaffold fibers resulting in a lack of hydrophilic groups (-COOH, -NH$_2$, =NH) on the fibers surface. On the other hand, complex heterogenic porous structure of the non-woven scaffold surface may limit its wettability [14]. Low concentration of L-arginine was confirmed by the absence of scaffold surface dyeing after treatment with ethanol solution of ninhydrin.

DSC curves of the fabricated PCL/L-arginine composite grafts are presented in figure 2:

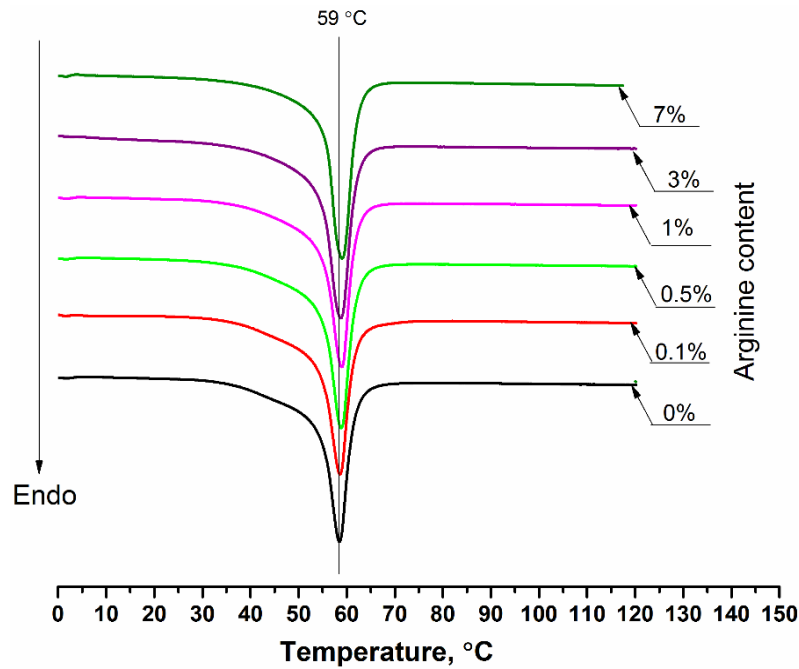

Figure 2. DSC curves of the obtained materials.

In studied temperature range, DSC curves of the obtained materials are presented by single endothermic peak at 59 °C, corresponding to melting temperature of PCL. With an increase of L-arginine content degree of crystallinity of the composite material was slightly changed (Table 1):

Table 1. Degree of crystallinity and mechanical properties of the obtained materials

|  | Degree of crystallinity, % | Elongation, % | Maximum strain, MPa | Young modulus, MPa |
|---|---|---|---|---|
| 0% | 53.0±0.5 | 162 ± 10 | 8 ± 1 | 18 ± 3 |
| 0.1% | 54.4±1.3 | 147 ± 13 | 12 ± 2 | 21 ± 3 |
| 0.5% | 55.5±1.6 | 123±12* | 10 ± 2 | 21 ± 3 |
| 1% | 53.7±0.3 | 142±7* | 14 ± 1* | 24 ± 2* |
| 3% | 56.4±0.5* | 151±10 | 17 ± 1* | 23 ± 1* |
| 7% | 58.2±0.4* | 176±23 | 21 ± 3* | 34 ± 3* |

\* - $p<0.05$ comparing to 0%

Effect of L-arginine content on elongation of the fabricated scaffolds is nonlinear. The least value of elongation was found at arginine content of 0.5%, the maximum one – at 7% (Table 1). It is possible that decrease of elongation of the composites with L-arginine concentrations up to 0.5-1% are connected with changing of fibers diameter distribution from bimodal to unimodal (Fig. 1). Further increase of elongation as well as maximum stress and Young modulus may be explained as follows. First, with an increase of L-arginine content, the amount of submicron fibers rises. That fact leads to decrease of extended defects concentration in the fiber, possibility of localization and further segregation of point defects in area comparable with nanofiber diameter and induces additional interactions between polymer molecules as a result of their orientation. Second, it was found that addition of L-arginine slightly enhances degree of crystallinity of the composites, thus improving their strength and Young modulus. Third, taking into account that COOH, $NH_2$, NH and =NH may form hydrogen bonds with ester -O-C(O)- fragments of PCL molecules, it may be supposed that PCL macromolecules may bind together via arginine molecules.

Results of biological studies of the fabricated scaffolds are presented in Table 2:

Table 2. Cell adhesion and viability

| Arginine content | Number of adhered cells, cells/mm$^2$ | Viable cells, % | Early Apoptosis, % | Late apoptosis and necrosis, % |
|---|---|---|---|---|
| 0% | 52 ± 16 | 66.8 ± 4.4 | 18.8 ± 5.3 | 12.3 ± 0.5 |
| 0.1% | 69 ± 27* | 67.2 ± 3.5 | 18.1 ± 0.1 | 12.5 ± 3.3 |
| 0.5% | 107 ± 32** | 71.6 ± 2.3 | 12.6 ± 2.2 | 14.9 ± 0.2 |
| 1% | 74 ± 25** | 76.4 ± 0.8 | 11.6 ± 0.4 | 11.0 ± 1.3 |
| 3% | 67 ± 24*** | 66.5 ± 8.6 | 16.9 ± 5.5 | 15.3 ± 2.8 |
| 7% | 113 ± 30** | 50.0 ± 7.7' | 27.6 ± 4.5 | 23.0 ± 0.1' |

\* $p=0.03$ – comparing to 0%,
\*\* $p<0.001$ – comparing to 0%,
\*\*\* $p=0.05$ – comparing to 0%,
' $p=0.02$ – comparing to 0%

Conducted studies demonstrate that in case of MMSC cultivation the optimal concentration of L-arginine is 0.5-1 % wt. as it provides the best cell adhesion and viability (Table 2). It may be suggested that these concentrations of L-arginine are enough to maintain needed and non-toxic level of nitric oxide (II).

**Conclusion**

Fibrous scaffolds made of poly(ε-caprolactone) and L-arginine were fabricated by electrospinning technique from their solution in hexafluoro-2-propanol. It was found, that an increase of arginine content in spinning solution leads to decrease of diameter of the formed fibers. With that, fibers diameter distribution changes from bimodal to unimodal and degree of the composite crystallinity, strength, and Young modulus increase. It was also observed that L-arginine content had no effect on scaffolds porosity and water contact angle. Concerning cytotoxicity results, the optimal concentration of L-arginine was found at 0.5-1 % wt. as it maintained the best cell adhesion and viability. As hexafluoro-2-propanol dissolves well wide range of biodegradable polyesters such as polyglycolide, polylactide, and their copolymers and mixtures, it is possible to develop a number of novel composite polymer materials containing L-arginine.

**Acknowledgements**

This research was funded by Russian Science Foundation (project№ 16-13-10239) and performed in Tomsk Polytechnic University and Almazov National Medical Research Centre.

546. doi:10.1039/tf9444000546.